\documentclass[conference, a4paper]{IEEEtran}
\usepackage[letterpaper, top=1.92cm, bottom=3.2cm, left=1.92cm, right=1.92cm]{geometry}
\IEEEoverridecommandlockouts

  	\usepackage[pdftex]{graphicx}
  	\graphicspath{{../pdf/}{../jpeg/}}
	\DeclareGraphicsExtensions{.pdf,.jpeg,.png}
	\usepackage[cmex10]{amsmath}
	\usepackage{mathabx}
	\usepackage{algorithmic}
	\usepackage{array}
	\usepackage{mdwmath}
	\usepackage[numbers,sort&compress]{natbib} 
	\usepackage{multirow} 
    \usepackage{xcolor}
\usepackage{harpoon}

\usepackage{booktabs}
\usepackage{fancyhdr}
    \usepackage{algorithmic}
    \usepackage[linesnumbered,boxed,ruled,commentsnumbered]{algorithm2e}
    \usepackage{float}
    \usepackage{soul}
    \usepackage{subfigure}
	\usepackage{mdwtab}
	\usepackage{eqparbox}
	\usepackage{url}
	\usepackage{tikz,xcolor}
    \definecolor{lime}{HTML}{A6CE39}
    \DeclareRobustCommand{\orcidicon}{%
    \begin{tikzpicture}
    \draw[lime, fill=lime] (0,0) circle [radius=0.16] node[white] {{\fontfamily{qag}\selectfont \tiny ID}};    
    \draw[white, fill=white] (-0.0625,0.095) circle [radius=0.007];    \end{tikzpicture}
    \hspace{-2mm}}
    \foreach \x in {A, ..., Z}{%
    \expandafter\xdef\csname orcid\x\endcsname{\noexpand\href{https://orcid.org/\csname orcidauthor\x\endcsname}{\noexpand\orcidicon}}
    }
\begin{document}

\title{\LARGE Task-oriented and Semantics-aware Communications for \\ Augmented Reality}

\author{\authorblockN{Leave Author List blank for your IMS2013 Summary (initial) submission.\\ IMS2013 will be rigorously enforcing the new double-blind reviewing requirements.}
\authorblockA{\authorrefmark{1}Leave Affiliation List blank for your Summary (initial) submission}}

 \author{\authorblockN{Zhe Wang,  Yansha Deng\authorrefmark{1}}
 \authorblockA{Department of Engineering, King's College London, London, UK\\
 \{tylor.wang, yansha.deng\}@kcl.ac.uk }

 \thanks{This work was supported in part by UKRI under the UK government’s Horizon Europe funding guarantee (grant number 10061781), as part of the European Commission-funded collaborative project VERGE, under SNS JU program (grant number 101096034). This work is also a contribution by Project REASON, a UK Government funded project under the FONRC sponsored by the DSIT.}
 }

\maketitle

\begin{abstract}
Upon the advent of the emerging metaverse and its related applications in Augmented Reality (AR), the current bit-oriented network struggles to support real-time changes for the vast amount of associated information, creating a significant bottleneck in its development. To address the above problem, we present a novel task-oriented and semantics-aware communication framework for augmented reality (TSAR) to enhance communication efficiency and effectiveness significantly. We first present an analysis of traditional wireless AR point cloud communication framework, followed by a detailed summary of our proposed semantic information extraction within the end-to-end communication. Then, we detail the components of the TSAR framework, incorporating semantics extraction with deep learning, task-oriented base knowledge selection, and avatar pose recovery. Through rigorous experimentation, we demonstrate that our proposed TSAR framework considerably outperforms traditional point cloud communication framework, reducing wireless AR application transmission latency by 95.6\% and improving communication effectiveness in geometry and color aspects by up to 82.4\% and 20.4\%, respectively.
\end{abstract}

\IEEEoverridecommandlockouts
\begin{keywords}
 Metaverse, augmented reality, semantic communication, semantics extraction, point cloud.
\end{keywords}

\IEEEpeerreviewmaketitle


\section{\small{Introduction}}
The metaverse, as an expansive digital universe, is poised to revolutionize various aspects of individual's daily life, primarily through applications such as Augmented Reality (AR), Virtual Reality (VR), and other immersive technologies. These technologies are reshaping areas from virtual conferencing to online gaming, and thus garnering attention from both industry and academia \cite{pacchioni2020virtual}.
However, real-time transmission of complex data such as avatars and point clouds poses considerable challenges, especially for avatar-centered applications that demand high bandwidth and rapid interaction. In contrast to earlier video-related applications dealing with images and text, the data-intensive nature of AR applications necessitates numerous packet transmissions, thereby intensifying bandwidth demands \cite{wang2022survey}. In general, a swift response time of less than 15 ms is crucial, which is substantially lower than the delay tolerated in video communication \cite{van2020human}. Therefore, redesign the communication framework is pivotal to mitigating time delays and bandwidth limitations in AR applications.

To address the high bandwidth challenges in AR applications, semantic communication has been proposed, aiming to facilitate more efficient communication by prioritizing task-relevant data \cite{kountouris2021semantics}. Early studies in this domain have focused on extracting semantic content within traditional data like text and images \cite{jiang2022wireless}, using Age of Information (AoI) as a metric to assess the timeliness of data \cite{maatouk2022age}. Despite these advancements, the practical application of semantic communication within AR remains unexplored. Moreover, current AoI-based strategies often fail to account for the importance of data sufficiency within emerging AR datasets, revealing a notable research gap. Hence, it is crucial to devise new techniques that effectively integrate semantic communication in AR applications, considering not only the timeliness but also the relevance and sufficiency of the data.

Current AR research mainly uses Head-Mounted Displays (HMDs), focusing on avatar-based applications to reduce computational and transmission loads while ensuring user privacy \cite{fernandez2022life}. Social media platforms, such as TikTok and Instagram, also use avatars for AR effects. Interestingly, current research demonstrate that avatars don't hinder social behaviors and even speed up tasks in gaming \cite{pauw2021avatar}. Virtual fitness and games like Pokemon Go also employ avatars to boost interactions. However, the efficiency of avatar transmission isn't optimal, and bandwidth issues persist. Lagging AR experiences are frequent in weak signal areas, demonstrating the current AR communication model's limitations. Various data types, including avatar skeleton and point cloud, are explored for avatar representation in AR wireless communication \cite{van2020objective}. Besides, the lack of a universal standard for avatar transmission in AR indicates a research gap, emphasizing the need to develop task-focused and semantics-aware communication for avatar representation in wireless AR.

Inspired by the 3D keypoints extraction method presented in \cite{you2020keypointnet}, we propose a task-oriented and semantics-aware communication framework in AR (TSAR) for avatar-centric end-to-end communication. In contrast to traditional point cloud AR communication framework that rely solely on point cloud input, our proposed TSAR extracts and transmits only essential semantic information. The contributions of our research can be summarized as follows:

\begin{enumerate}
\item We propose a task-oriented and semantics-aware communication framework for augmented reality (TSAR) in an avatar-centric conferencing and gaming AR application,
which significantly enhances the efficiency and effectiveness of wireless communication compared to the traditional point cloud communication framework.

\item We innovatively initiate rigorous ablation experiments and define base knowledge and semantic information within our proposed TSAR framework, specifically for avatar-centric conferencing and interactive gaming AR applications for bolstering communication efficiency and enriching the AR experience.

\item 
Compared with the point cloud communication framework, our proposed TSAR framework achieves better client side viewing in terms of color quality, geometry quality, and transmission delay with improvements of up to 20.4\%, 82.4\%, and 95.6\%, respectively. 
\end{enumerate}

The rest of the paper is organized as follows: Section II presents the system model and problem formation, encompassing both the traditional point cloud and the TSAR framework. Section III elaborates on the proposed methods for both semantics and task-level optimization, including semantics extraction with deep learning, task-oriented base knowledge selection, and avatar pose recovery. Section IV outlines the evaluation indicators and experiment performance. Finally, Section V concludes the paper.

\section{\small{System Model and Problem Formation}}
\begin{figure*}[t]
  \centering
  \includegraphics[width=0.9\textwidth ]{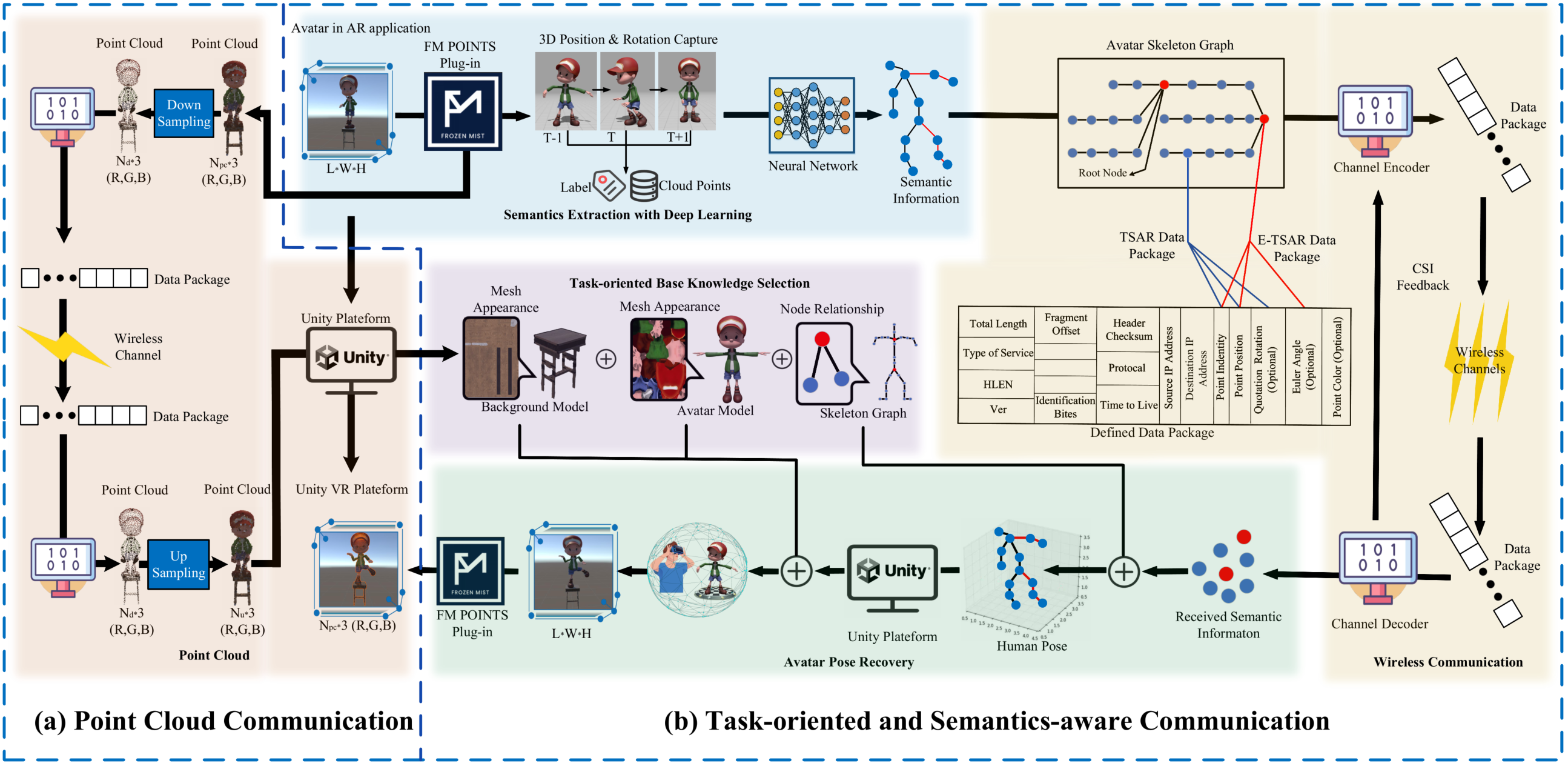} 
  \caption{Wireless communication frameworks: (a) Point cloud communication framework    (b) Task-oriented semantics-aware communication framework} 
  \label{wirelesscommunication} 
\end{figure*}

In this section, we first present the traditional point cloud communication framework for AR applications. Then, we subsequently introduce our novel TSAR framework that incorporates both semantics and task levels, as depicted in Fig. \ref{wirelesscommunication}. Finally, we present the problem formation associated with our proposed framework.
\subsection{System Model}
\subsubsection {Traditional Point Cloud Communication Framework}
We focus on an avatar-centric AR application in conferencing and gaming, which is the prominent scenery in the metaverse \cite{fernandez2022life}. These applications need real-time transmissions of avatar and background model for HMD display in a spatial volume defined by dimensions $L$, $H$, and $W$. For a smooth AR experience, high-resolution point clouds of both avatar and background models are generated, compression and transmitted. Utilizing plugins on the Unity3D platform, such as FM POINTS, allows real-time point cloud transformation of AR scenery. The generated point clouds $\mathbf{P}_\text{ar}$ consist of numerous points $\overrightharp{v}_i$, which could be written as $\mathbf{P_{\text{ar}}}=\left[\left.\overrightharp{v}_i\right|_0 ^{N_\text{pc}}\right]^{\text{T}}$, where ${N}_{\text{pc}}$ is the total number of point clouds. Typically, over 1,500 thousand point clouds per frame is needed to represent each 3D object for a satisfactory QoS \cite{zhang2021towards}, with each point contains location and color information.
These point clouds then follow a transmission process of point cloud communication framework, as represented in the Fig. \ref{wirelesscommunication} (a). The procedure commences with point cloud downsampling to reduce the volume of data, followed by wireless transmission. Upon reception, point cloud upsampling is performed to restore the full detail of the 3D objects. Finally, these upsampled point clouds are displayed on the Unity3D platform.

\subsubsection {Semantic Information Extraction}
Unlike the traditional point cloud communication, which heavily relies on raw data sensing and acquisition, our proposed TSAR framework incorporates a comprehensive method that simultaneously processes and compresses point cloud data at both task and semantics levels. This approach significantly reduces the data size since only the semantics and task-related data are extracted and transmitted.
The process begins with point cloud sensing data, $\mathbf{P}_\text{ar}$, which encapsulates all models within the AR scenery. Here, only the moving avatar's position, requiring a refresh in the subsequent frame, is considered crucial. As a result, the semantics extraction process yields the skeleton information of the avatar, which is crucial for avatar pose representation.

To effectively recover the avatar pose at the client side, the skeleton information, $\overrightharp{I}_i$, needs to encompass both the avatar model's position and quaternion rotation \cite{gonzalez2020movebox}, which can be symbolically expressed as
\begin{equation}
\label{skeletoninfor}
\overrightharp{I}_i=(l_\text{x}, l_\text{y}, l_\text{z},r_\text{x}, r_\text{y}, r_\text{z}, r_\text{w}) \ i \in [1,N_\text{a}],
\end{equation}
where, $N_\text{a}$ denotes the total number of avatar skeletons, $\overrightharp{l}_i=(l_\text{x}, l_\text{y}, l_\text{z})$ represents a three-dimensional location, and $\overrightharp{r}_i=(r_\text{x}, r_\text{y}, r_\text{z}, r_\text{w})$ denotes the quaternion rotation. The semantics extraction process, $\mathcal{S}(\cdot)$, is defined as
\begin{equation}
\mathbf{D}_{\text{tsar}}= \mathcal{S}(\mathbf{D}_{\text{pc}}, {\theta}_\text{s})= [{\overrightharp{I}_\text{1}, \overrightharp{I}_\text{2},...,\overrightharp{I}_{{N}_{\text{a}}}}]^\text{T},
\end{equation}
where ${\theta}_\text{s}$ represents all the neural network and setup parameters utilized in the semantics extraction, and $\mathbf{D}_{\text{tsar}}$ symbolizes the entire semantic information encapsulated in the moving avatar's skeleton $\overrightharp{I}_i$.

\subsection{Wireless Channel Model}
Our wireless communication model is characterized by a Rayleigh fading channel, impacted by additive white Gaussian noise and utilizing an Orthogonal Frequency-division Multiplexing (OFDM) scheme. The OFDM approach divides the physical channel into multiple parallel subchannels. Each subchannel experiences varying levels of noise, leading to different Signal-to-Noise Ratios (SNRs).

Before wireless transmission, Binary Phase-shift Keying (BPSK), a widely adopted modulation technique, transforms analog signals into digital bits. This can be done by modifying the phase of a carrier signal in response to the data values inputted into the system. Once BPSK processing is complete, the resulting bits, denoted as $s_n$, are ready for transmission.
The multi-path channel within the OFDM scheme can be described as ${\overrightharp{H}_{\text{c}}}=\left(\left.h_n\right|_0 ^{N_\text{c}}\right)$
where $N_\text{c}$ stands for the total number of subchannels in $\overrightharp{H}_c$, and $h_n$ signifies the channel gain of the $n$-{th} subchannel.

Considering the characteristics of each subchannel, the cumulative SNR of the communication process within channel $\overrightharp{H}_\text{c}$ is expressed as
\begin{equation}
\label{snrequ}
\text{SNR}=\frac{\sum_{n=1}^{N_\text{c}}\left\|h_n \cdot s_n\right\|^2}{\sum_{n=1}^{N_\text{c}} \sigma_{n}^2},
\end{equation}
where, $\sigma_{n}^{2}$ represents the noise within the $n$-{th} subchannel. Various semantic information is transmitted through different subchannels in the OFDM wireless channel. 

\subsection {Problem Formation}
In summary, the overall TSAR framework aims to achieve efficient data transmission for better avatar representation in wireless AR applications and improve the client's QoS through semantics extraction. The primary objective of the framework is to minimize the transmitted data package in wireless communication and maximize the client-side AR viewing experience based on the transmitted semantic information. The objective function of the entire TSAR framework can be expressed as
\begin{equation}
\begin{aligned}
& \mathcal{P}: \min _{(\theta_\text{s})} \lim _{T \rightarrow+\infty} \frac{1}{T} \sum_{t=0}^T \sum_{i=0}^{N_\text{a}}\left(\overrightharp{I}_i^t-{\overrightharp{I}}_i^{t'}\right)
\ \ \  i \in[1, N_\text{a}],
\end{aligned}
\end{equation}
where $\overrightharp{I}_i^t$ represents the semantic information of the skeleton at time $t$, and ${\overrightharp{I}}_i^{t'}$ is the received semantic information after the wireless channel, which is detailed in Eq. (\ref{skeletoninfor}). This equation formulates the problem of minimizing the error in avatar representation during wireless communication.

\section{Proposed Methods}
In this section, we will demonstrate the design principles of TSAR optimization at the semantics and task levels in detail. 
\subsection{Semantics Extraction with Deep Learning}
To efficiently extract semantic information from point cloud, we design a semantics-aware network named SANet to extract the skeleton keypoint information of a moving avatar from the point clouds of AR scenery. 
The SANet operates by using point cloud data ${\mathbf{D}_{\text{pc}}}$ as input, which represents the 3D coordinates of both the background environment and the moving avatar. This data is then processed by the SANet to extract accurate the avatar skeleton information, crucial for faithfully reproducing the avatar's movements and interactions in the virtual environment.
The design objective of the SANet is to minimize the Euclidean distance ($\mathcal{L}2$) between the semantic information according to predicted skeleton index, denoted as $\mathcal{S}({\mathbf{D}_{\text{pc}}})$, and the labeled semantic information of the skeleton location, represented as $\mathbf{D}^{l}_{\text{tsar}}$. The interplay between these variables is defined as
\begin{equation} \text{Loss}=\arg \min \underset{\left(\theta_{\text{s}}\right.)}{\mathcal{L}2}\left(\mathcal{S}({\mathbf{D}_{\text{pc}}}), \mathbf{D}^{l}_{\text{tsar}}\right), \label{Losstheta} \end{equation}
where the symbol $\theta_{\text{s}}$ stands for all the neural network parameters in the SANet. Training the SANet entails optimizing these neural network parameters in $\theta_{\text{s}}$ to minimize the $\mathcal{L}2$ loss.


\subsection{Task-oriented Base Knowledge Selection}
In contrast to traditional point cloud wireless communication framework, the TSAR performs the avatar pose recovery differently with the transmission of the base knowledge $\boldsymbol{B}$ at the beginning of AR application. As illustrated in the base knowledge part in Fig. \ref{wirelesscommunication}, the base knowledge encompasses different types of information: avatar skeleton graph $\mathcal{G}$, moving avatar model, avatar location $l_0$, avatar model $\mathcal{A}_o$, stationary background model $\mathcal{S}_o$, and their respective appearance meshes, $\mathcal{M}_{a}$ and $\mathcal{M}_{s}$. Whenever a new model appears in the AR scene, the base knowledge at both transmitter and receiver need to be updated synchronously.

Apart from quaternion rotation, current research also employs euler angles to represent rotations in AR scenery. In comparison to quaternion, euler angles offer a simpler and more information-efficient method for representing rotation and calculating root node position when skeleton graph $\mathcal{G}$ is available to represent node connection. Euler angles define pitch $e_\text{p}$, roll $e_\text{y}$, and yaw $e_\text{r}$ to represent rotations around the three primary axes with an associated root point. This approach necessitates less information to reconstruct the avatar's skeleton pose compared to quaternion \cite{quintero2022excite}, resulting in smaller data packets and potentially more efficient communication. The transformation from rotation to euler angles can be expressed as
\begin{equation}
\label{eulerangle}
{ 
\left[\begin{array}{c}
{e_\text{p}} \\
{e_\text{r}} \\
{e_\text{y}}
\end{array}\right]=\left[\begin{array}{c}
\arctan \large \frac{2\left(r_\text{y} r_\text{z}+r_\text{w} r_\text{x}\right)}{1-2\left(r_\text{x}^2+r_\text{y}^2\right)} \\
\arcsin \left(2\left(r_\text{w} r_\text{y}-r_\text{x} r_\text{z}\right)\right) \\
\arctan \frac{2\left(r_\text{x} r_\text{y}+r_\text{w} r_\text{z}\right)}{1-2\left(r_\text{y}^2+r_\text{z}^2\right)}
\end{array}\right] *\frac{180}{\pi}.
}
\end{equation}

To better explore the most suitable base knowledge, we have designed the following ablation experiments for semantic communication with different shared base knowledge and semantic information definitions\footnote{Semantic information, as shown in Fig. \ref{wirelesscommunication}, consists of the skeleton information that need to be transmitted in every frame. Conversely, base knowledge encompasses information used primarily in the first frame.}, which include the base TSAR framework (TSAR) and Euler angle based TSAR framework (E-TSAR).

\textbf{TSAR}: 
In the base TSAR framework, semantic information for each skeleton is defined as the data pertaining to position and quaternion rotation, as illustrated in Eq. (\ref{skeletoninfor}). The shared base knowledge comprises the background model, moving avatar model, and their corresponding appearance meshes, which could be represented as
\begin{equation}
\boldsymbol{B}=\{\mathcal{A}_o, \mathcal{S}_o, \mathcal{M}_a, \mathcal{M}_s\}.
\end{equation}

\textbf{E-TSAR}: 
Based on the TSAR, the semantic information in each skeleton is defined as the position and euler angle in E-TSAR, according to Eq. (\ref{eulerangle}) which could be defined as
\begin{equation}
\label{semanticinforationetasr}
\overrightharp{I}^{e}_i = {(e_\text{r}, e_\text{y}, e_\text{p})}, \ i \in [1,{N}_{\text{a}}],
\end{equation}
where $\overrightharp{I}^{e}_i={(e_\text{r}, e_\text{y}, e_\text{p})}$ denotes the euler angle. The shared base knowledge $\boldsymbol{B}^\text{e}$ of E-TSAR encompasses the avatar skeleton graph, shared background model, moving avatar model, avatar location $l_0$, and their appearance meshes, which is defined as 
\begin{equation}
\label{baseknowledgedaiG}
\boldsymbol{B}^\text{e}=\{\mathcal{A}_o, \mathcal{S}_o, \mathcal{M}_a, \mathcal{M}_s, l_0, \mathcal{G}\}.
\end{equation}

\begin{algorithm}[t]
	\renewcommand{\algorithmicrequire}{\textbf{Input:}}
	\renewcommand{\algorithmicensure}{\textbf{Output:}}
	\caption{Avatar Pose Recovery}
        \label{humanposerecovery}
	\begin{algorithmic}[1]
		\STATE Initialization: Base knowledge $\boldsymbol{B}$, received data $\textbf{D}_{\text{tsar}}$
            \STATE Get skeleton graph $\mathcal{G}$ and avatar model $\mathcal{A}_o$ from $\boldsymbol{B}$
            \STATE Initialize a graph $\mathcal{G}'$ with the same format of $\mathcal{G}$
		\STATE  Count skeleton number $N_\text{a} = \mathcal{C}_\text{s}(\mathcal{G}) $
  		\STATE  Count received data each frame $N_\text{r} = \mathcal{C}_\text{r}(\mathbf{D}_{\text{tsar}}) $
            \IF{ (${\mathcal{G} \notin \boldsymbol{B}} \And {l_i \in \mathbf{D}_{\text{tsar}}})$ }
                \FOR{each $i$ in $N_\text{r}$}
                \STATE Attach $\overrightharp{I}_i$ to graph $\mathcal{G}'$
                \ENDFOR
            \ELSE
                \FOR{each $i$ in $N_{a}$}
                \STATE update $l_i$ according to Eq. (\ref{eq_l11}) and Eq. (\ref{eq_l12})
                \STATE Attach $\overrightharp{I}^{e}_i$ to graph $\mathcal{G}'$
                \ENDFOR
            \ENDIF
            \STATE Generate avatar $\hat{\mathcal{A}}_o$ with model $\mathcal{A}_o$, appearance mesh $\mathcal{M}_a$, and skeleton information graph $\mathcal{G}'$
		\ENSURE  Avatar with recovered position $\hat{\mathcal{A}}_o$
	\end{algorithmic}  
\end{algorithm}

\begin{figure*}[th]
  \centering
  \includegraphics[width=0.8\textwidth]{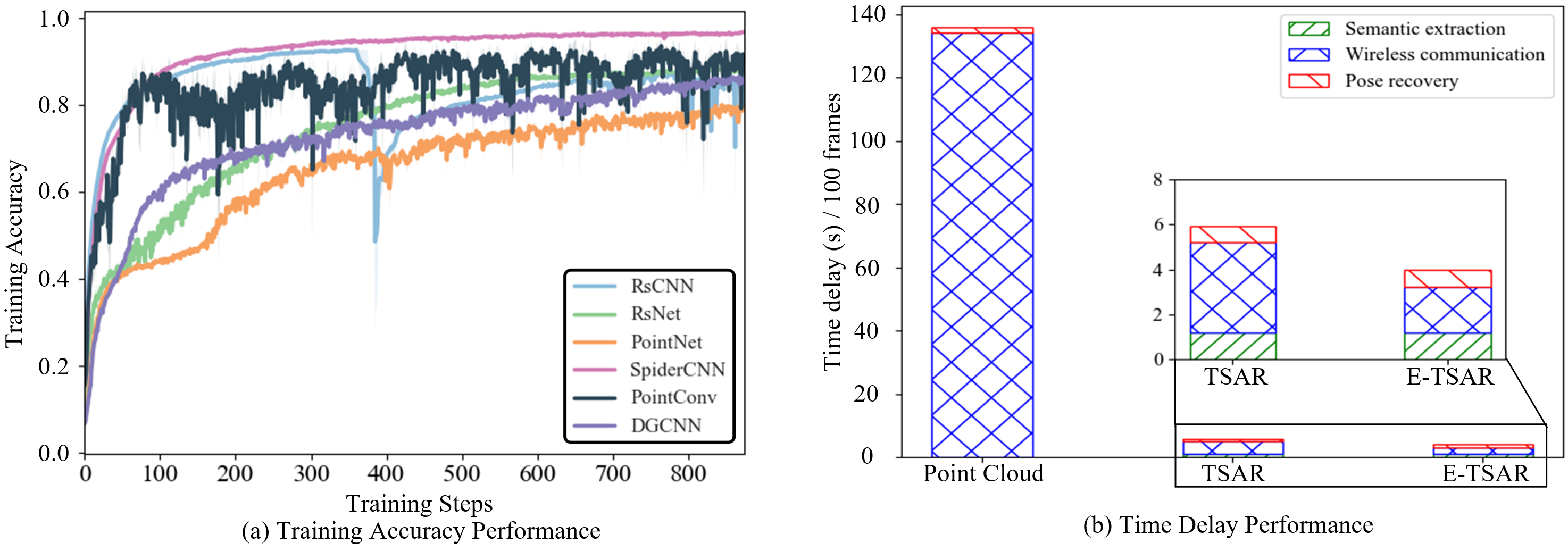} 
  \caption{Performance of SANet training accuracy and communication time delay} 
  \label{experiment1} 
\end{figure*}

\subsection{Avatar Pose Recovery}
The details of the avatar pose recovery involve using the skeleton graph $\mathcal{G}$ in the base knowledge and the received semantic information to reconstruct the avatar position. The entire avatar position recovery process is shown in Alg. \ref{humanposerecovery}. Specifically, a recursive algorithm is employed to traverse and assign all skeleton information to the avatar skeleton graph with initialized parameters. However, due to differences in the definition of the semantic information and the shared base knowledge, the avatar poses recovery process has variations between the base TSAR and E-TSAR framework.

On the one hand, the base TSAR framework employs a simple avatar pose recovery method, assigning the avatar model with value based on the skeleton point identity using the received position and quaternion rotation in the semantic information. On the other hand, the E-TSAR framework, which only transmits the euler angle of each skeleton point as semantic information, 
reconstructs the avatar pose by first determining the relationships between the skeleton points in the $\mathcal{G}$. It then computes the position of each skeleton point by considering its euler angle, avatar location $l_0$, and the position of its root skeleton within the $\mathcal{G}$. The relative distance $\Delta l_{(i,i-1)}$ between skeleton node $\overrightharp{I}_i$ and its root node $\overrightharp{I}_{i-1}$, which can be represented as
\begin{equation}
\label{eq_l11}
\Delta l_{(i,i-1)} = \overrightharp{r}_i \times l_{i-1},
\end{equation}
where $r_i$ represents the eular angle of skeleton node $\overrightharp{I}_{i}$, and the actual position of skeleton node $\overrightharp{I}_i$ will be calculated by combining $\Delta l_{(i,i-1)}$ and the root position of $l_{i-1}$, which can be expressed as
\begin{equation}
\label{eq_l12}
l_i= l_{i-1} + \Delta l_{(i,i-1)},
\end{equation}
where $l_i$ represents the position of the $i$-{th} skeleton node in the avatar skeleton graph $\mathcal{G}$.

\section{Experiments}
In this section, we compare our proposed TSAR with traditional point cloud framework and conduct numerous experiments to verify the superiority of the proposed TSAR framework.

\subsection{Point Cloud Communication Framework}
Adapting the method from \cite{fujihashi2021holocast+}, we propose a baseline point cloud framework named the Point Cloud. The framework, as depicted in Fig. \ref{wirelesscommunication} (a), incorporates point cloud downsampling with the farthest point sampling algorithm and upsampling with linear interpolation algorithm. 

\subsection{Semantics Extraction}
To determine the effectiveness of the designed SANet, and achieve outstanding performance, we train the SANet with various backbone networks, including ResNet, RsCNN, PointNet, SpiderCNN, PointConv, and DGCNN. Similar to \cite{you2020keypointnet}, we use the mean Average Precision (mAP) as the performance evaluation metric to assess the prediction accuracy of the predicted keypoint probabilities in relation to the ground truth semantic information labels. Fig. \ref{experiment1} (a) demonstrates that the SpiderCNN-based SANet attains the highest accuracy in skeleton information extraction, exceeding the accuracy of over 96\% among the same training epochs. This result demonstrates the SANet achieves significantly more reliable and stable prediction compared to other backbone-based networks.

\begin{figure*}[t]
  \centering
  \includegraphics[width=0.95\textwidth]{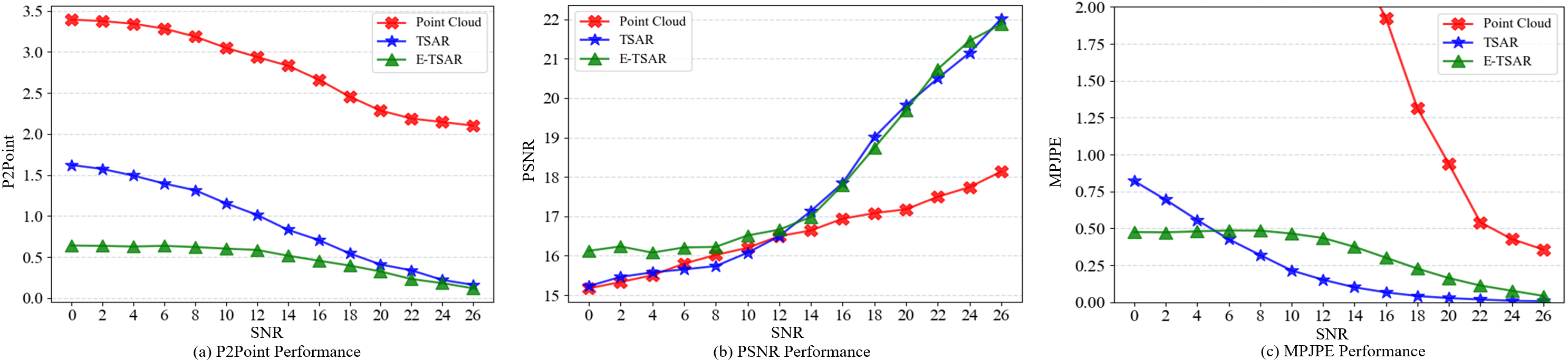} 
  \caption{Performance of P2Point, $\text{PSNR}_\text{y}$, and MPJPE} 
  \label{experiment2} 
\end{figure*}

\subsection{Evaluation Indicators}
To thoroughly evaluate the performance of our TSAR framework, we implement metrics to verify the communication efficiency and effectiveness in AR application. These metrics include Point-to-Point (P2Point), Peak Signal-to-Noise Ratio of luminance component ($\text{PSNR}_\text{y}$), Mean Per Joint Position Error (MPJPE), and transmission delay. 

\textbf{P2Point}: The P2Point metric is employed to assess the alignment of point cloud data between the transmitter $\mathbf{P}_\text{t}$ and receiver $\mathbf{P}_\text{r}$. P2Point quantifies the geometric discrepancy relative to the transmitted and received points acts as a benchmark for gauging the quality of the point cloud.

$\textbf{PSNR}_\textbf{y}$: The $\text{PSNR}_\textbf{y}$ metric facilitates the evaluation of the luminance component of the point cloud. It works by mapping the color of each point in the $\mathbf{P}_\text{t}$ to the nearest color in the $\mathbf{P}_\text{r}$ and computing the PSNR for the luminance component error.

\textbf{MPJPE}: The MPJPE metric is used to estimate human poses, serving as a tool to evaluate the discrepancy in avatar pose between the transmitter and receiver.

\textbf{Transmission Delay}: The transmission delay metric is used to estimate and ensure service quality. The delay in the entire TSAR comprises different components, such as semantics extraction, wireless communication, and avatar pose recovery.

\subsection{Performance Evaluation}

Fig. \ref{experiment1} (b) illustrates the significant reduction in transmission delay achieved by our TSAR and E-TSAR frameworks in comparison to the traditional point cloud framework. Despite the inclusion of an additional semantics extraction step, the TSAR and E-TSAR frameworks manage to curtail transmission time by approximately 95.6\% and 97\% compared with point cloud framework, respectively. This remarkable efficiency is due to the TSAR framework transmitting a mere 25 skeleton points, as opposed to the 2,048 points required by point cloud communication. This considerable reduction in transmission packages significantly cuts down bandwidth usage. These impressive outcomes underscore the viability of our TSAR in facilitating high bandwidth AR applications in dynamic wireless environments.

Fig. \ref{experiment2} (a) and Fig. \ref{experiment2} (b) plot the P2Point and $\text{PSNR}_\text{y}$ results, respectively. The P2Point metric assesses geometric discrepancies throughout the complete AR scene, not limited to the avatar or any specific model. As the SNR increases, the P2Point results is ordered as E-TSAR$\textless$TSAR$\textless$Point Cloud, and the pattern becomes more pronounced with SNR decrease. Specifically, the E-STAR decrease the P2Point for about 82.4\% compared with point cloud communication, indicating significant distortion in avatar positions when using the point cloud framework.
Conversely, the $\text{PSNR}_\text{y}$ metric measures the differences in the color aspect between the transmitter and the receiver. Here, both TSAR and E-TSAR surpass the traditional point cloud framework by over 20.4\% at the best SNR scenery. With the $\text{PSNR}_\text{y}$ value ordering as E-TSAR$\textgreater$TSAR$\textgreater$Point Cloud, it is evident that TSAR and E-TSAR, due to their ability to create compact and grouped point clouds, demonstrate their ability in minimizing distortion effectively. This demonstrates TSAR's high effectiveness in preserving color quality and its potential to improve wireless AR applications.

Fig. \ref{experiment2} (c) plots the MPJPE results which measure the variance in avatar skeleton position between receiver and transmitter. The TSAR and E-TSAR notably surpasses the point cloud communication framework across all SNR scenarios. Particularly, TSAR achieves higher avatar pose recovery in higher SNR scenarios with an 83.3\% decrease in MPJPE compared to point cloud framework, highlighting its superior effectiveness in avatar transmission. However, in lower SNR conditions, E-TSAR outperforms TSAR in MPJPE difference, suggesting that utilizing shared avatar model as base knowledge helps in preserving movements within the avatar's capabilities, thereby enhancing AR avatar representation from distortion.



\section{Conclusion}
This paper has proposed a task-oriented and semantics-aware communication framework for Augmented Reality that aims to achieve reliable and low-latency wireless communication in the AR application. By defining the semantic information and base knowledge in the AR, our proposed TSAR have successfully 
reducing wireless AR application transmission latency by 95.6\% and improving communication effectiveness 
by up to 82.4\% and 20.4\%, respectively.
Our future work includes exploring the scalability of our framework to support larger virtual environments and investigating the integration of other semantic features to improve the accuracy and efficiency of communication.


\footnotesize
\bibliographystyle{ieeetr}
\bibliography{biblio_traps_dynamics}

\end{document}